\documentclass[apjl,a4paper,12pt,useAMS]{emulateapj}
\usepackage{amssymb}
\usepackage{graphicx}

\newcommand{\be}{\begin{eqnarray}}
\newcommand{\ee}{\end{eqnarray}}
\newcommand{\el}{ et al.}

\begin{document}
\title{Compact Binary Mergers and The Event Rate of Fast Radio Bursts}
\author{Xiao-Feng Cao$^{1}$, Yun-Wei Yu$^{2,3}$, and Xia Zhou$^{4}$}
\altaffiltext{1}{School of Physics and Electronics Information,
Hubei University of Education, Wuhan 430205, China}
\altaffiltext{2}{Institute of Astrophysics, Central China Normal
University, Wuhan 430079, China, {yuyw@mail.ccnu.edu.cn}}
\altaffiltext{3}{Key Laboratory of Quark and Lepton Physics (Central
China Normal University), Ministry of Education, Wuhan 430079,
China}
\altaffiltext{4}{Xinjiang Astronomical Observatory, Chinese Academy of Sciences, Urumqi 830011, China}

\begin{abstract}
Fast radio bursts (FRBs) are usually suggested to be associated with
mergers of compact binaries consisting of white dwarfs (WDs),
neutron stars (NSs), or black holes (BHs). We test these models by
fitting the observational distributions in both redshift and
isotropic energy of 22 Parkes FRBs, where,  as usual, the rates of
compact binary mergers (CBMs) are connected with cosmic star
formation rates by a power-law distributed time delay. It is found
that the observational distributions can well be produced by the CBM
model with a characteristic delay time from several ten to several
hundred Myr and an energy function index
$1.2\lesssim\gamma\lesssim1.7$, where a tentative fixed spectral index $\beta=0.8$ is adopted for all FRBs. Correspondingly, the local event
rate of FRBs is constrained to $(3-6)\times10^4f_{\rm b}^{-1}(\mathcal T/270\rm
s)^{-1}(\mathcal A/2\pi)^{-1}\rm ~Gpc^{-3}yr^{-1}$ for an adopted
minimum FRB energy of $E_{\min}=3\times10^{39}$ erg, where $f_{\rm b}$ is the beaming factor of the radiation, $\mathcal
T$ is the duration of each pointing observation, and $\mathcal A $ is
the sky area of the survey. This event rate, about an order of
magnitude higher than the rates of NS-NS/NS-BH mergers, indicates
that the most promising origin of FRBs in the CBM scenario could be
mergers of WD-WD binaries. Here a massive WD could be produced since
no FRB was found to be associated with a type Ia supernova.
Alternatively, if actually all FRBs can repeat on a timescale
much longer than the period of current observations, then they could
also originate from a young active NS that forms from relatively
rare NS-NS mergers and accretion-induced collapses of WD-WD
binaries.
\end{abstract}
\keywords{radio continuum: general --- stars: neutron --- stars:
white dwarfs}

\section{Introduction}
Studies on mergers of binary systems composing of a pair of compact
objects, i.e., white dwarfs (WDs), neutron stars (NSs), or black
holes (BHs), are of fundamental importance in astrophysics, because
these mergers have or might have tight connections with current and
future detections of gravitational waves \citep{Abbott2016,Abbott2017},
with the formation of heavy elements via r-process \citep{Lattimer1974,Eichler1989,Hotokezaka2013,Bauswein2013,Just2015}, with the production of type Ia supernovae \citep[SNe Ia;][]{Tutukov1981,Webbink1984}, and with the origin of short gamma-ray bursts \citep[GRBs;][]{Paczynski1986,Eichler1989,Guetta2006,Coward2012} as well as mergernova/kilonova emission \citep{Li1998,Metzger2010,Yu2013,Coulter2017,Evans2017}.  The detection of various possible
electromagnetic radiation from compact binary mergers (CBMs) can
play a crucial role in uncovering the nature of progenitor binaries
and in locating and identifying the associated gravitational wave
signals.

Recently, it was suggested that some CBMs, specifically, mergers of
double WDs \citep{Kashiyama2013}, of double NSs \citep{Totani2013,Wang2016,Yamasaki2017}, of a NS and a BH \citep{Mingarelli2015}, or even of two charged BHs \citep{Zhang2016}, could be
responsible for the newly-discovered fast radio bursts (FRBs). FRBs
are millisecond radio transients of intensities of a few to a few
tens of Jansky at $\sim 1$ GHz \citep{Lorimer2007,Keane2012,Keane2016,Thornton2013,Burke-Spolaor2014,Spitler2014,Ravi2015,Masui2015,Champion2016,Caleb2017,Petroff2017,Bannister2017}. Due to the short durations of FRBs and the low angular
resolution of present radio surveys, it is difficult to capture
counterparts of FRBs in other wavelength bands, even if these
couterparts indeed exist. This makes it impossible to directly
determine the distances of FRBs\footnote{For the only repeated FRB,
FRB 121102, its host galaxy and a persistent radio counterpart have
been detected and then its redshift has been measured to $z=0.19$,
which undeniably confirmed its cosmological origin \citep{Chatterjee2017,Marcote2017,Tendulkar2017}.} and to
identify their origins. In any case, the anomalously high dispersion
measures (DMs; $\sim 200-2600\rm ~pc~cm^{-3}$) of FRBs, which are
too high to be accounted for by the high-latitude inter-stellar
medium in the Milky Way, robustly suggest that the FRBs could have
cosmological distances of redshifts up to $z\sim 4.0$. Therefore,
the isotropically equivalent energy release of an FRB can be
estimated to within the range of $\sim10^{39-42} \rm erg$. In the
suggested CBM models, such an energy could be naturally provided by
the inspiral of the binary or the spin-down of the remnant object
due to magnetic dipole radiation and magnetospheric activities.
Furthermore, it is believed that this energy should be released via
coherent radiations, with some similarity to the pulse radiation of
pulsars \citep{Yang2017a}.

Besides the energy scale and time scale of FRBs, another crucial
constraint on models is the event rate of FRBs and, furthermore,
its redshift-dependence. During the past several years, the
increasing FRB number has already enabled statistical
investigations of FRBs \citep{Yu2014,Bera2016,Caleb2016,Li2017,Katz2016a,Oppermann2016,Lu2016,Vedantham2016,Fialkov2017,Lawrence2017,Cao2017a,Cao2018,Macquart2018}. In particular, \cite{Cao2017a} found that the proportional coefficient between the FRB rates and cosmic star formation rates (CSFRs) could be redshift-dependent, which somewhat favors the CBM model. Therefore, in this paper we confront the CBM model with the number
distributions of FRBs in redshift as well as in energy (see \cite{Yamasaki2017} for a relevant calculation). By fitting the
observational distributions, we test the feasibility of the CBM
model for explaining the FRB phenomena and, simultaneously,
constrain the model parameters. According to observational
constraints, the possible nature of the progenitor compact binaries
can be discussed.

%---------------------------------------------------------------------------------------------------------
%---------------------------------------------------------------------------------------------------------
\section{The model}
\subsection{The rate of CBMs}

A merger takes place after a compact binary loses their orbital
energy through gravitational radiation.  The rate of CBMs at
redshift $z$ can be related to the CSFR
at redshift $z'$ that is determined by the time delay equation as
$t(z')=t(z)-\tau$, where $t(z)=\int_z^{\infty}[(1+z')H(z')]^{-1}dz'$
is the age of the universe at redshift $z$ and $H(z)\equiv
H_0\sqrt{\Omega_{\rm M}(1+z)^3+\Omega_{\Lambda}}$. Hereafter the
cosmological parameters are taken as $\Omega_M=0.32$,
$\Omega_\Lambda=0.68$, and $H_0=70\rm km~s^{-1}Mpc^{-1}$. The delay time $\tau$ is determined by both
the gravitational radiation decay of the binary orbit and the
formation process of the compact binary. The latter factor is
further related to the supernova mechanism, the natal kick velocity
of NS, the mass transfer between the binary stars, and etc \citep{Portegies Zwart1998,Belczynski2002,Mennekens2010,Chruslinska2018}. Considering of a
probability distribution of $P(\tau)$ of the delay times, the
rate of CBMs can be calculated by the following convolution \citep{Piran1992,Guetta2006}:
\begin{eqnarray}
\dot{R}_{\mathrm{m}}(z)&\propto&\int_{0}^{t(z)-t(z_{\rm b})}\dot{\rho}_{*}[t(z)-\tau]P(\tau)d\tau\nonumber\\
&\propto&\int^{z_{\rm b}}_{z}\dot{\rho}_{*}(z')P[t(z)-t(z')]{dt\over dz'}dz'\label{CBMR},
\end{eqnarray}
where $\dot{\rho}_{*}(z)$ is the CSFR and $dt/dz=-[(1+z)H(z)]^{-1}$.
The upper limit of the above integrate, $z_{\rm b}$, is set at the
redshift when the binaries start forming.

Following a series of measurements of CSFRs, a consensus on the
history of cosmic star formation emerges up to redshift $z\sim4$
\citep{Hopkins2006}. In the Hopkins \& Beacom's data, a trend
of decrease of the CSFRs appears in higher redshift range. This
trend was further confirmed by the observations of Lyman break or
Ly-$\alpha$ emitter galaxies \citep{Bouwens2012,Bouwens2015,Oesch2013,Oesch2014,Coe2012,McLeod2016} and long GRBs
\citep{Chary2007,Yuksel2008,Kistler2009,Wang2009,Ishida2011,Tan2015}, although
there is still a debate on the decrease rate at high redshifts. The
high-redshift CSFRs can also be constrained by the Gunn-Peterson
trough observations to quasars and by the Thomson scattering optical
depth of cosmic microwave background photons \citep[e.g.,][]{Yu2012,Wang2013}. Combining the various measurements and constraints, we
take the cosmic star formation history as follows \citep{Yu2012}:
%(Porciani \& Madau 2001)
%\begin{eqnarray}
%\dot{\rho}_*(z)=\dot{\rho}_*(0){23e^{3.4z}\left[\Omega_m(1+z)^3+\Omega_\Lambda\right]^{1/2}\over(e^{3.4z}+22)(1+z)^{3/2}},\label{SFR}
%\end{eqnarray}
\begin{equation}
\dot{\rho}_*(z)\propto\left\{
   \begin{array}{cc}
   (1+z)^{3.44},&{~\rm for ~}z<0.97,\\
   (1+z)^{0},&{~\rm for ~}0.97\leq z<3.5,\\
   (1+z)^{-0.8},&{~\rm for ~}z\geq 3.5,
      \end{array}
\right.
\end{equation}
with a local CSFR of $\dot{\rho}_*(0)=0.02\rm
M_{\odot}~yr^{-1}Mpc^{-3}$. In any case, the uncertainty of the
high-redshift CSFRs would not significantly influence the rates of
CBMs at relatively low redshifts of most FRBs.

The lifetime of gravitational radiation decay of a binary orbit is
determined by the initial orbital separation ($a_{i}$) and the
initial ellipticity ($e_i$).  According to the relation $\tau\propto
a_{i}^4$ and assuming $P(a_{i})\propto a_{i}^{q}$, \cite{Piran1992}
suggested
\begin{equation}
P(\tau)= P(a_{i}) {da_{i}\over d\tau}\propto\tau^{(q-3)/4},
\end{equation}
where the initial ellipticity is taken as a constant. A reference
value of $q=-1$ can further be inferred from the data of regular
binaries, which yields\begin{equation}
P(\tau)\propto1/\tau.\label{TD0}
\end{equation}
On the one hand, this simple power-law distribution has been
generally confirmed by more elaborate calculations \citep{Greggio2005,Belczynski2006,Mennekens2010,Ruiter2011,Mennekens2016}. This indicates that delay times are dominated by the gravitational radiation. On the other
hand, such a delay time distribution has been widely and
successfully applied in modeling the redshift distribution of SNe
Ia originating from mergers of double WDs \citep{Totani2008,Maoz2012} and of short GRBs originating from mergers
of double NSs or a NS and a BH \citep{Guetta2006,Nakar2006,Virgili2011,Hao2013,Wanderman2015}. Furthermore, this power-law distribution could also be
supported by the observations of six double NS systems \citep{Champion2004}.

\begin{figure}
\centering\resizebox{0.48\textwidth}{!} {\includegraphics{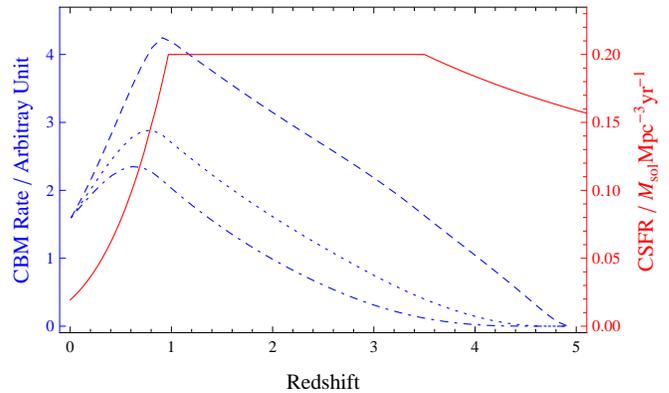}}
\caption{CBM rate as a function of redshift for different
characteristic delay times: $\tau_{\rm c}=$100, 500, and 1000 Myr
(dashed, dotted, and dash-dotted lines, respectively), where $z_{\rm
b}=5$ is taken. The solid line represents the adopted star formation
history. } \label{fig_CBMR}
\end{figure}

The delay time distribution is usually found to be peaked at a
cutoff value, $\tau_{\rm c}$, below which the probability decreases
drastically. Therefore, we tentatively take an empirical expression
as follows:
\begin{equation}
P(\tau)\propto\left({\tau\over\tau_{\rm c}}\right)^{-1}e^{-\tau_{\rm c}/\tau},\label{TD}
\end{equation}
with which we derive the CBM rate as a function of redshift from Eq.
(\ref{CBMR}). The result is primarily dependent on the  value of the
crucial cutoff of the delay times, as presented in Figure
\ref{fig_CBMR}. Numerical simulations show that the value of
$\tau_{\rm c}$ is probably around a few hundred Myr for double NS
mergers, but around a few ten Myr for NS-BH mergers \citep{Mennekens2016,Chruslinska2018}. The delay time
distribution of SNe Ia can usually be described by a broken-power
law consisting of $\tau^{-0.5}$ and $\tau^{-1}$ \citep{Greggio2005,Graur2013}, where the former power law is probably determined by
the formation time of WDs. Therefore, the value of $\tau_{\rm c}$
for double WD mergers can in principle be defined by the break time
between the two power laws, which also ranges from several ten to
several hundred Myr.

\subsection{Model-predicted FRB numbers}
It is assumed that a particular type of CBMs produce the observed FRBs of isotropic energy releases of $E$, which could satisfy a power-law distribution as
\begin{eqnarray}
\Phi(E)\equiv{dN\over dE}\propto{E}^{-\gamma}, {\rm ~for~}E\geq
E_{\min},
\end{eqnarray}
where the value of $E_{\min}$ can roughly be inferred from
observations. The combination of the above intrinsic energy
distribution with the observational thresholds of telescopes
determines the fraction of FRBs that can be detected by the
telescopes. For a specific telescope survey, the observational
number of FRBs in the redshift range $(z_1,z_2)$ or in the energy
range $(E_1,E_2)$ can be calculated by
\begin{eqnarray}
N&=&\mathcal T{\mathcal A \over 4\pi}f_{\rm b}
\int_{z_1}^{z_2}\dot{R}_{\rm m}(z){dV(z)\over 1+z}\left[\int_{\max[E_{\rm th}(z),E_{\min}]}^{E_{\max}}\Phi(E)dE\right],\nonumber\\\label{number2}
\end{eqnarray}
or
\begin{eqnarray}
N&=&\mathcal T{\mathcal A \over 4\pi}f_{\rm b}
{\int_{E_{1}}^{E_2}}\Phi(E)\left[\int_{0}^{\min[z_{\rm h}(E),z_{\max}]}\dot{R}_{\rm m}(z){dV(z)\over 1+z}\right]dE,\nonumber\\\label{number3}
\end{eqnarray}
where $\mathcal T$  is the duration of each pointing observation,
$\mathcal A $ is the sky area of the survey, $f_{\rm b}$ is the beaming factor of the FRB radiation, $dV(z)=4\pi d_c(z)^2
cH(z)^{-1}dz$ is the comoving volume element,
$d_c(z)=c\int_0^{z}H(z')^{-1}dz'$ is the comoving distance, and the
factor $(1+z)$ represents the cosmological time dilation for the
observed rates. The energy threshold of a telescope involved in Equations (\ref{number2}) and (\ref{number3}) can be determined by
\begin{eqnarray}
E_{\rm th}(z)&=&4\pi d_c(z)^2(1+z)\Delta\nu\mathcal F_{\rm \nu,th}
k(z) , \label{Ethz}
\end{eqnarray}
where $\Delta \nu$ and $\mathcal F_{\rm \nu,th}$ are the frequency
bandwidth and the fluence sensitivity of the telescope,
respectively. The correction factor $k(z)$ converts the FRB energy
from the observational band ($\nu_1,\nu_2$) into a common emitting
frequency range ($\nu_a,\nu_b$) for all FRBs. By assuming a
power-law spectrum, $\mathcal F_{\nu}\propto\nu^{-\beta}$, the
$k-$correction can be calculated to
\begin{eqnarray}
k(z)=\frac{{\nu_b}^{(1-\beta)}-{\nu_a}^{(1-\beta)}}{{{[(1+z)\nu_2]}^{(1-\beta)}-{[(1+z)\nu_1]}^{(1-\beta)}}
}. \label{kz}
\end{eqnarray}
Finally, the horizon redshift $z_{\rm h}(E)$ appearing in integral
(\ref{number3}) can be solved from the equation $E=E_{\rm th}(z_{\rm
h})$, which means that, for an isotropic energy of FRBs, the
observational horizon of the telescope is at $z_{\rm h}$. The
maximum redshift $z_{\max}$ corresponds to the maximum DM below
which the FRB searches were conducted. One must keep in mind that a
remarkable number of FRBs of relatively high redshifts and of
relatively low energies have be missed by the present telescope
surveys due to the telescope thresholds, when the observational
distributions of FRBs are discussed and used.

%
%------------------------------------------------------------------------
%------------------------------------------------------------------------
%

%
%------------------------------------------------------------------------
\begin{figure}
\centering\resizebox{0.45\textwidth}{!} {\includegraphics{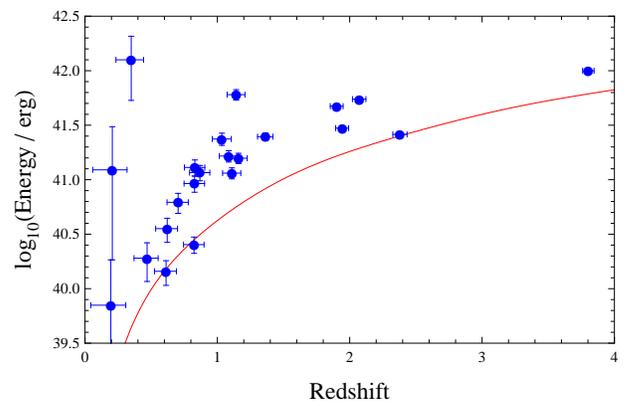}}
\caption{The 22 Parkes FRBs in the $z-E$ plane. The errors of the
data correspond to the uncertain range of DM$_{\rm host}$ from zero
to $200\rm~pc~cm^{-3}$ and the central values are given for DM$_{\rm
host}=100\rm~pc~cm^{-3}$. The FRB energies are corrected for a tentative spectral index $\beta=0.8$. The solid line represents the
observational energy threshold of the Parkes telescope below which
the identification opportunity of an FRB decreases drastically.
}\label{data}
\end{figure}
%------------------------------------------------------------------------
%

%
%------------------------------------------------------------------------
\begin{figure*}
\centering\resizebox{0.45\textwidth}{!}
{\includegraphics{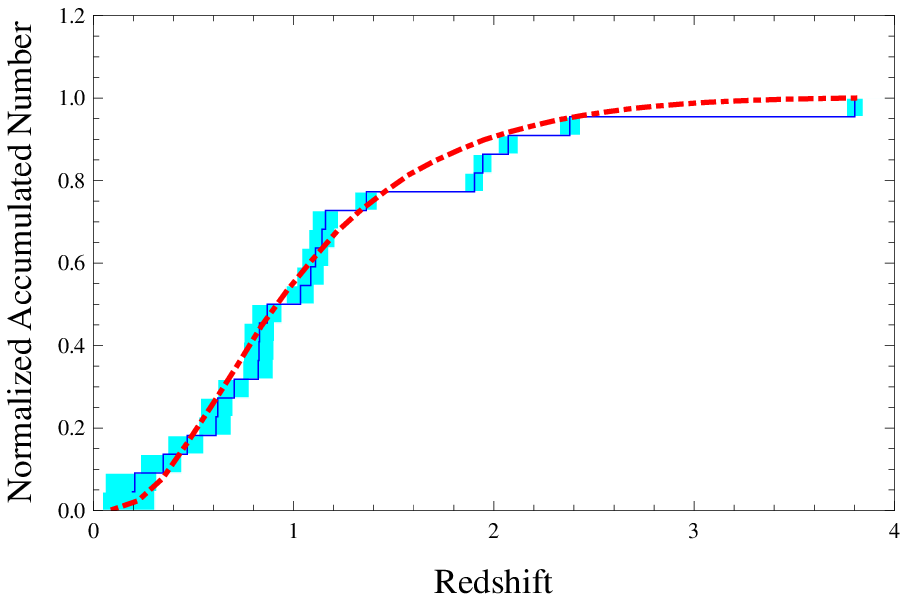}}\resizebox{0.45\textwidth}{!}
{\includegraphics{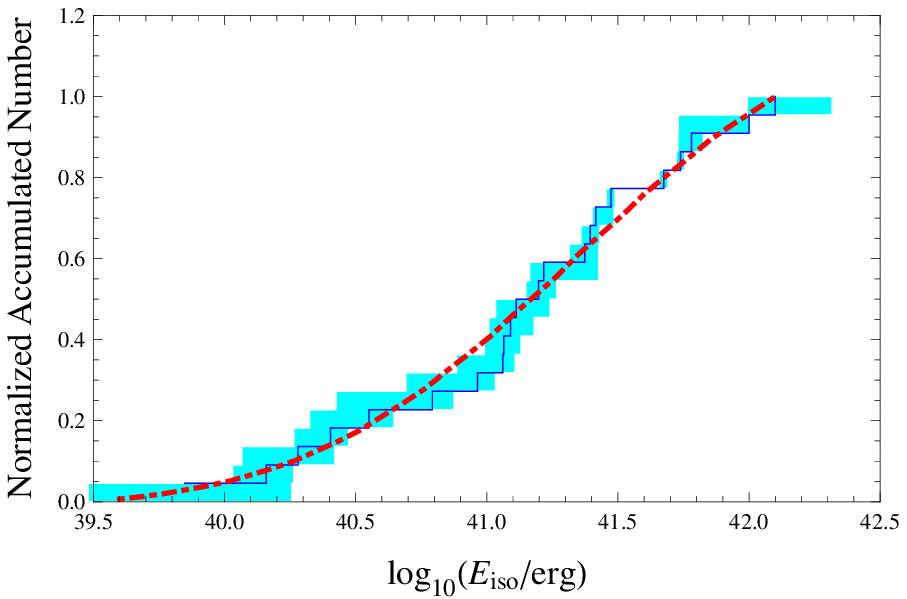}} \caption{Normalized accumulated
distributions in redshift (left) and isotropic energy (right) of the
22 Parkes FRBs. The solid line is obtained by fixing DM$_{\rm host}$ to $100\rm~pc~cm^{-3}$ for all FRBs, while the shadow represents the uncertainty of the FRB distribution arising from the variation of DM$_{\rm host}$ from $0\rm~pc~cm^{-3}$ to $200\rm~pc~cm^{-3}$. An
example fitting by the CBM model to the FRB distribution for DM$_{\rm host}=100\rm~pc~cm^{-3}$ is presented by the dash-dotted line, where the model parameters are taken as $\beta=0.8$, $\gamma=1.4$, and $\tau_{\rm c}=350$ Myr.}\label{fig_best}
\end{figure*}
%------------------------------------------------------------------------
%

%
%------------------------------------------------------------------------
\begin{figure*}
\centering\resizebox{0.3\textwidth}{!} {\includegraphics{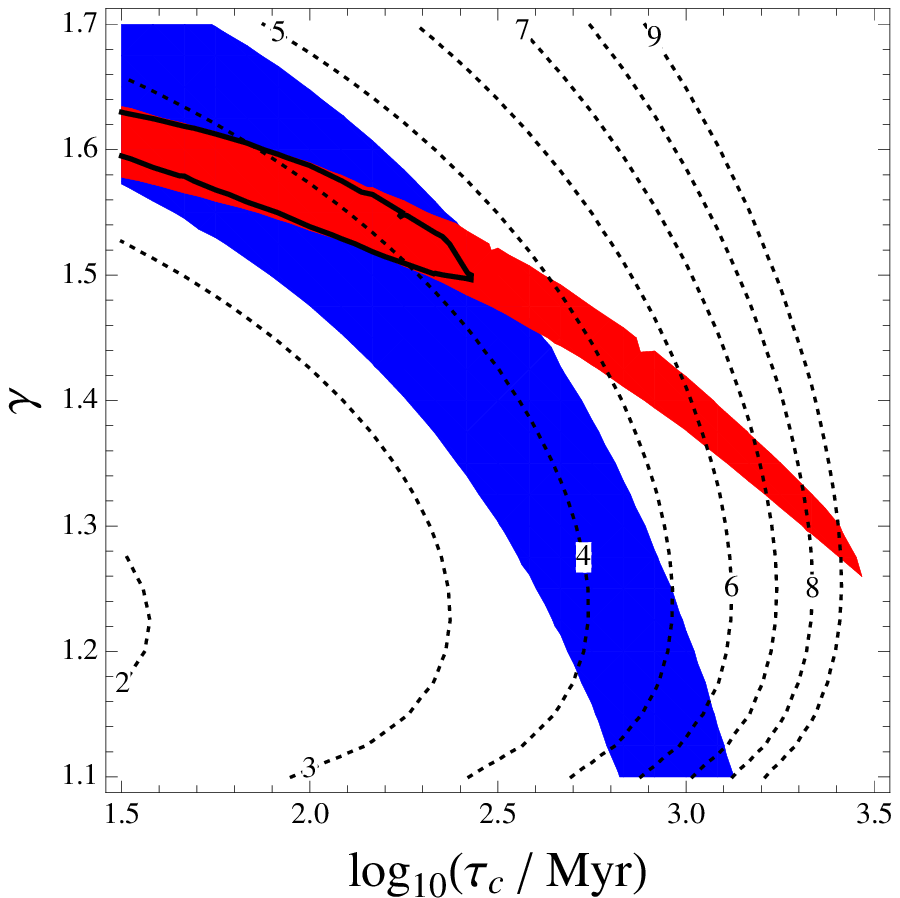}}
\resizebox{0.3\textwidth}{!}
{\includegraphics{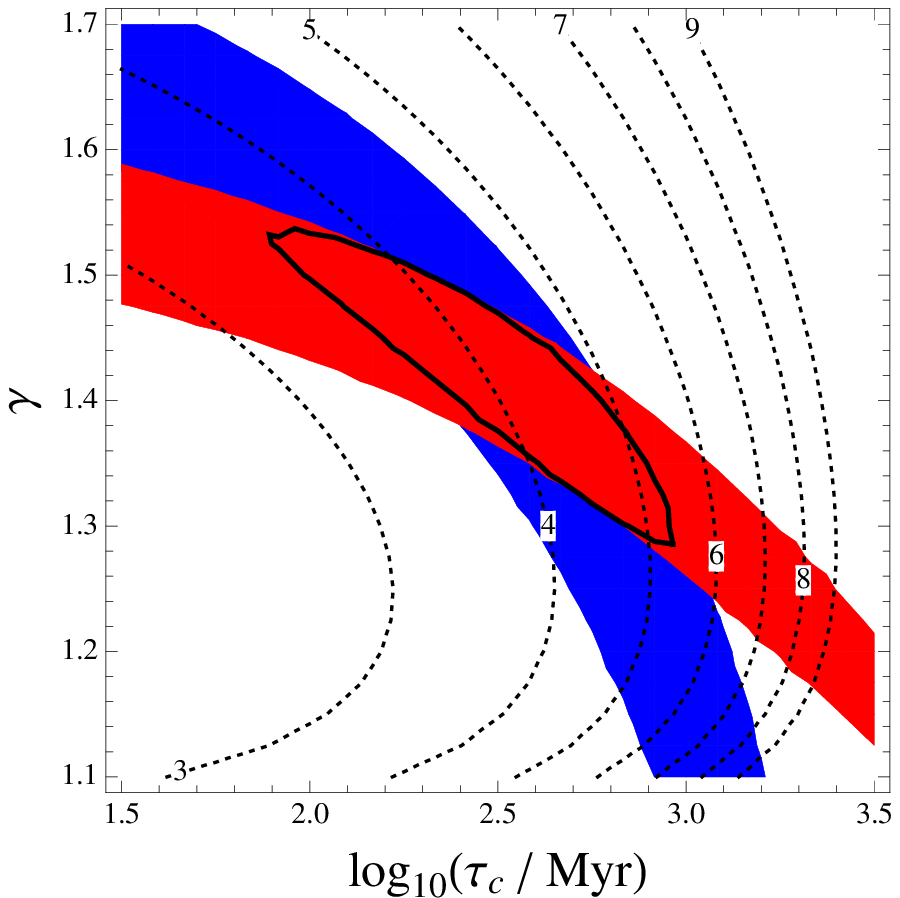}}\resizebox{0.3\textwidth}{!}
{\includegraphics{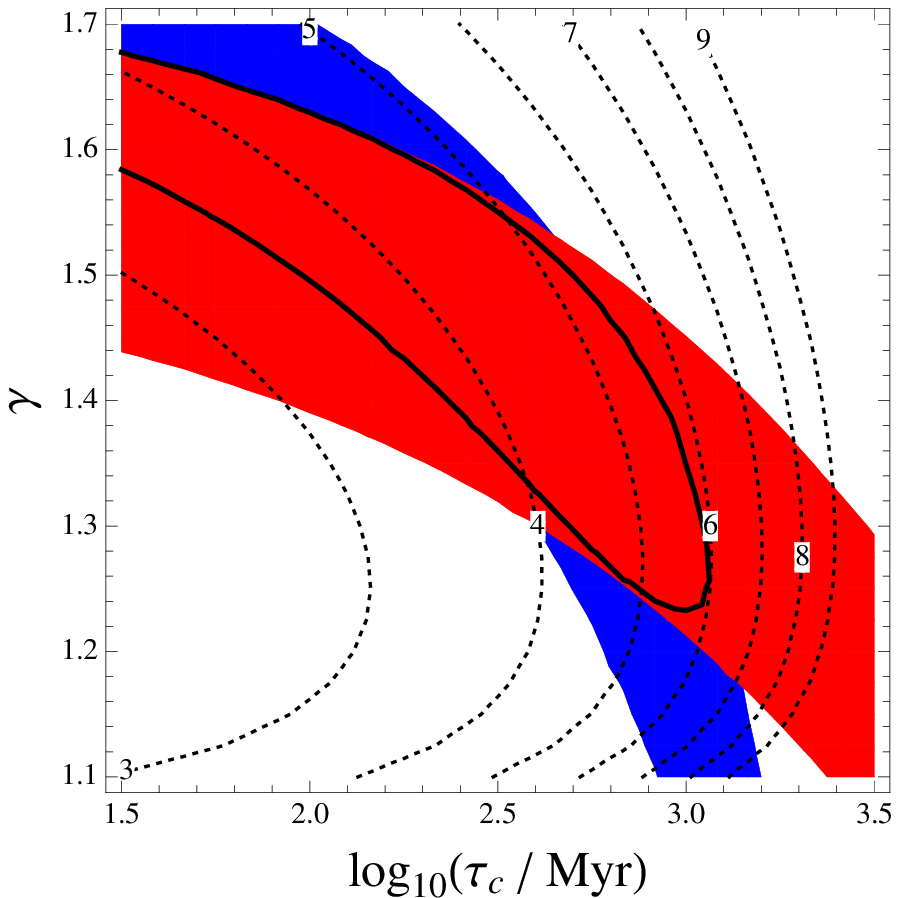}}\caption{95\% confidence level contours
of parameters $\tau_{\rm c}$ and $\gamma$ given by the
Kolmogorov-Smirnov test. The blue and red contours correspond to
the fittings of redshift and energy distributions, respectively. The
overlapped region of the contours is presented by the solid line.
The dashed lines indicate the required local event rate of FRBs as
labeled by $\dot{R}_{\rm m}(0)/10^4\rm Gpc^{-3}yr^{-1}$. From left
to right, the value of DM$_{\rm host}$ is fixed to $\rm 0~pc~cm^{-3}
$, $\rm 100~pc~cm^{-3} $, and $\rm 200~pc~cm^{-3}$,
respectively}.\label{fig_probability}
\end{figure*}
%------------------------------------------------------------------------
%

%---------------------------------------------------------------------------------------------------------
%---------------------------------------------------------------------------------------------------------
\section{Fitting to observational distributions}
Up to FRB 171209, a total of 30 FRBs have been detected by different
telescopes including the Parkes, UTMOST, GBT, ASKAP, and Arecibo,
which are cataloged on the website {\it http://frbcat.org/} (see
\cite{Petroff2016} and references therein). In this paper we only
take into account the largest sub-sample provided by the Parkes
containing 22 FRBs, so that the very different parameters of
different telescopes will not be involved. The only repeated FRB
121102 discovered by Arecibo is just excluded.  The redshifts of
the FRBs can be inferred from their DMs, by subtracting the
contributions from the Milky Way and the host galaxies, while the DM
of the FRB sources are considered to be relatively much lower.
Specifically, the following equation is used to calculate the
redshifts of the FRBs \citep{Ioka2003,Inoue2003,Yang2017b}:
\begin{eqnarray}
{\rm DM}_{\rm IGM}(z)&=&{\rm DM_{}-DM_{MW}}-{{\rm DM}_{\rm
host}\over{ 1+z}}\nonumber\\
&=&f_{\rm IGM}f_{\rm e}{3cH_0\Omega_{\rm b}\over 8\pi G m_{\rm
p}}\int_0^z{H_0(1+{z'})\over {H(z')}
}dz',
\end{eqnarray}
where $f_{\rm IGM}\sim0.83$ is the fraction of baryon mass in
intergalactic medium (IGM), $f_{\rm e}\sim7/8$ is the number ratio
between free electrons and baryons (including proton and neutron) in
IGM, $m_{\rm p}$ is the proton mass, and $\Omega_{\rm b}=0.04$.
While the values of DM$_{\rm MW}$ have been provided in the catalog,
the DM$_{\rm host}$ are completely unknown. In any case, a rough
estimation on the order of magnitude of DM$_{\rm host}$ could still
be made by referring to the observation of the host galaxy of FRB
121102, which gives\footnote{The DM of FRB 121102 contributed by its
host galaxy was suggested by \cite{Tendulkar2017} to
\begin{eqnarray}
{\rm DM_{host,FRB 121102}\thickapprox 324 ~pc~cm^{-3}}\left[{4d_{\rm
kpc}f\over\zeta(1+\epsilon^2)}\right]^{1/2},\label{DMhost}
\end{eqnarray}
where $d_{\rm kpc}=d/\rm kpc$ is the total path length of the FRB
emission through the galactic disk, $f$ is the faction of the path
that is occupied by ionized clouds, $\zeta\geq1$ defines
cloud-to-cloud density variations in the ionized regions,
$\epsilon\leq1$ is the fractional variation inside discrete clouds
due to turbulent-like density variations.} DM$_{\rm host}\sim100\rm
~pc~cm^{-3}$, although it is not clear whether or not this repeated
FRB has an origin identical to the non-repeated ones. Therefore, in
our calculations we take the values of DM$_{\rm host}$ varying from
zero to $\rm 200~pc~cm^{-3}$, where the upper bound is set according
to the present minimum DM of the Parkes FRBs, i.e., DM$_{\rm FRB
150807}=266.5\rm ~pc~cm^{-3}$. This variation range of DM$\rm
_{host}$ leads to the uncertainty of redshifts and energies of the FRBs, as
shown in Figure \ref{data}.

With an inferred redshift, the isotropically equivalent energy of an
FRB can be calculated by
\begin{equation}
E=4\pi d_c(z)^2(1+z) \Delta\nu \mathcal F_\nu k(z),
\end{equation}
where the Parkes parameters for $k-$correction are taken as follows:
$\nu_1=1182$ MHz and $\nu_2=1522$ MHz (i.e., $\Delta \nu=0.34$ GHz
centering at $1.35$ GHz) and then $\nu_a=1182$ MHz and $\nu_b=7610$
MHz which correspond to the redshift range of $0\lesssim z\lesssim4$
of observed FRBs. For simplicity, a tentative spectral index $\beta=0.8$ is assumed
in view of our very poor knowledge of the FRBs' spectra and the high degeneracy between $\beta$ and $\gamma$ due to the $k$-correction. Two Parkes
FRBs have published spectral indices including FRB 131104 with
$\beta=0.3\pm0.9$ \citep{Ravi2015} and FRB 150418 with
$\beta=1.3\pm0.5$ \citep{Keane2016}. However, it should be
cautioned that these values are very sensitive to the true position
of the FRBs within the telescope beam pattern. Finally, the energy
threshold of the Parkes telescope, presented by the solid line in
Figure \ref{data}, is calculated by Equation (\ref{Ethz}) with a
fluence sensitivity as \citep{Bera2016,Caleb2016}
\begin{equation}
\mathcal F_{\rm \nu,th}=0.04{\rm S\over N}\left[{\Delta t_{\rm
obs}(z)\over1\rm ms}\right]^{1/2}\rm Jy ~ms,
\end{equation}
where the characteristic minimum signal-to-noise ratio is adopted to
$\rm S/N=10$ and the typical FRB duration $\Delta t_{\rm obs}(z)$ as
a function of redshift is given by fitting the observational
duration distribution as did in \cite{Cao2017a}.

In view of the limit number of observed FRBs, we only pay attention
to the accumulated distributions of the 22 Parkes FRBs,
as presented in Figure \ref{fig_best}. These FRB distributions are somewhat uncertain due to the uncertainty of DM$_{\rm host}$. For a fixed DM$_{\rm host}$, we
can fit the observational distributions by Equations (\ref{number2})
and (\ref{number3}) by varying the values of
the most crucial model parameters, i.e., $\tau_{\rm c}$ and $\gamma$. The goodness of the fits is assessed by the
Kolmogorov-Smirnov test, where the observational uncertainties are not involved. Then, for a general consideration, we carry out such calculations for three different values of DM$_{\rm
host}$, i.e., $0\rm ~pc~cm^{-3}$, $100\rm ~pc~cm^{-3}$, and $200\rm ~pc~cm^{-3}$. As a result, the $95\%$ confidence level regions of parameters $\tau_{\rm c}$ and $\gamma$ are presented in Figure \ref{fig_probability} by two contours deriving from the fittings of the redshift and energy distributions, respectively. The large overlap of the two contours demonstrates that sufficiently good fits of observations
can easily be found in the CBM model. One example of the best fits to the distributions for DM$_{\rm
host}=100\rm ~pc~cm^{-3}$ is showed by the
dash-dotted line in Figure \ref{fig_best}, which is given by $\tau_{\rm c}=350$ Myr, $\gamma=1.4$, and $E_{\min}=3\times10^{39}$ erg. The results for different values of DM$_{\rm
host}$ together indicate that, while $1.2\lesssim\gamma\lesssim1.7$, the characteristic delay time $\tau_{\rm c}$ can range from several ten to several hundred Myr\footnote{If we release the fixing of the value of $\beta$, our constraints on the model parameters can be somewhat changed, in particular, for the parameter $\gamma$ \citep{Cao2017a} because of its tight connection with $\beta$ through the $k$-correction of FRB energies. However, the value of $\tau_{\rm c}$ would not substantially deviate from the large range presented here.}, which is broadly
consistent with the theoretical expectations of the CBM model. This somewhat favors the CBM explanation of the FRB phenomena, although the range of $\tau_{\rm c}$ is still too large to fix the nature of the compact binaries.

Finally, by using the total FRB number of 22, we can determine the
local rate of the FRB-related CBMs for different values of
$\tau_{\rm c}$ and $\gamma$, as labeled by the dashed lines in
Figure \ref{fig_probability}. According to the overlapped regions of
the contours for all different DM$_{\rm host}$ cases, we can have
\begin{equation}
\dot{R}_{\rm m}(0)\approx (3-6)\times10^4{\rm Gpc^{-3}yr^{-1}}f_{\rm b}^{-1}\left({\rm
\mathcal T\over 270\rm s}\right)^{-1}\left({\rm \mathcal A\over
2\pi}\right)^{-1},\label{mergerrate}
\end{equation}
where the reference values of $\mathcal T$ and $\mathcal A$ are
taken by referring to \cite{Thornton2013}. Substituting the
above local event rate into Equation (\ref{CBMR}) and integrating
$\dot{R}_{\rm m}(z)$ from $z=0$ to $4$, we can obtain the full-sky
event rates for different fluence sensitivities, as listed in Table
1, where $\tau_{\rm c}=350$Myr, $\gamma=1.4$, and $\dot{R}_{\rm
m}(0)=4.1\times10^4\rm Gpc^{-3}yr^{-1}$ are taken. For the
sensitivity $\mathcal F_{\nu,\rm th}=0.4$ Jy ms corresponding to the
Parkes, the presented rate of 14,080 $\rm day^{-1}sky^{-1}$ can be
easily understood by the following calculation:
\begin{eqnarray}
\dot{R}_{\rm FRB,full-sky}&=&{1\over f_{\rm b}}\cdot{N_{\rm FRB,Parkes}\over \mathcal
T}\cdot{4\pi \over
\mathcal A} \nonumber\\
&=&14,080~{\rm day^{-1}sky^{-1}}f_{\rm b}^{-1}
\left({\rm
\mathcal T\over 270\rm s}\right)^{-1}\left({\rm \mathcal A\over
2\pi}\right)^{-1}.
\end{eqnarray}

\begin{table}
\caption{Full-sky FRB event rates for different sensitivities}
\begin{center}
\begin{tabular}{|c|c|}
\hline
$\mathcal F_{\nu,\rm th}$ for $\Delta t_{\rm obs}=1$ ms &Event rate\\
(Jy ms) &  (Number/day/sky)\\
\hline
0.2&20,000\\
0.4&14,080\\
1.0&8,100\\
3.0&3,500\\
\hline
\end{tabular}
\end{center}
\label{default}
\end{table}%

%---------------------------------------------------------------------------------------------------------
%---------------------------------------------------------------------------------------------------------
\section{Conclusion and discussions}
The fitting results presented in this paper indicate that the CBM
model with reasonable parameter values can well account for the FRB
phenomena in the sight of the redshift dependence of event rate, although the uncertainty of model parameters is still large. It is at least indicated that the FRB rates could be connected with CSFRs by power-law distributed delay times and the FRB energy distribution could be effectively expressed by a single power law. Furthermore, the relatively certain value of the local event rate of FRBs enables us to discuss the nature of the
compact binaries, specifically, two WDs, two NSs, or a NS and a BH.

Mergers of NS-NS and NS-BH binairies are long considered to be
progenitors of short GRBs, which was recently confirmed by the
discovery of GRB 170817A and the associated gravitational wave event
GW 170817. On the one hand, according to GW 170817, the rate of
NS-NS mergers has been directly inferred to $\dot{R}_{\rm
ns-ns}(0)\sim1540^{+3200}_{-1220}\rm Gpc^{-3}yr^{-1}$ \citep{Abbott2017}. An absolute upper limit on this rate was previously imposed
to 12,600 $\rm Gpc^{-3}yr^{-1}$ by the non-detection of this type of
mergers during O1 of LIGO \citep{Abbott2016}. On the other hand,
during the past decade, the local event rate of short GRBs has been
widely investigated and found to be from a few to a few ten $\rm
Gpc^{-3}yr^{-1}$ \citep{Guetta2006,Nakar2006,Guetta2009,Dietz2011,Coward2012,Wanderman2015,Tan2018,Zhang2018}. According to the latest statistics, we can
get $\dot{R}_{\rm sGRB}(0)\approx4\rm Gpc^{-3}yr^{-1}$ for an
assumed minimum luminosity of short GRBs of
$L_{\min}\sim5\times10^{49}\rm erg~s^{-1}$. The conversion of this
short GRB event rate to merger rate is highly dependent on the
measurements of openning angles of GRB jets. For a possible range of
the angles of $5^{\circ}-30^{\circ}$, the local merger rate can be
inferred to $\dot{R}_{\rm ns-ns}(0)\sim(30-1100)\rm
Gpc^{-3}yr^{-1}$, which is broadly in agreement with the LIGO
result. Meanwhile, the rate of NS-BH mergers is considered to be
comparable to or more probably lower than the rate of NS-NS
mergers \citep{Abadie2010}. Therefore, it seems difficult to explain all FRBs by
only NS-NS and NS-BH mergers \citep[cf.][]{Callister2016}.

Mergers of double WDs could lead to different outcomes
including SN Ia explosions, a stable WD, and a stable NS through
accretion-induced collapse (AIC) \citep{Canal1976,Nomoto1991}. Simulations showed that the local rate of WD mergers can reach
several times $ 10^4 \rm Gpc^{-3} yr^{-1}$ \citep{Badenes2012},
which was supported by the measurement of SN Ia rate as
$(3.01 \pm 0.062) \times 10^4 \rm Gpc^{-3} yr^{-1}$ \citep{Li2011}
although SNe Ia can also originate from a single WD accreting from its campanion star. The
general consistency between the WD merger rate and the rate
presented in Equation (\ref{mergerrate}), if the beaming of the FRB radiation can be ignored, indicates that the WD
mergers could be the most promising origin of FRBs in the CBM scenario. So far
there was no bright SNe Ia reported to be associated with observed
FRBs. Therefore, the plausible origin of FRBs is the
formation of a massive WD as suggested by
\cite{Kashiyama2013} or a stable AIC NS. Here, the fraction of AICs of WD-WD mergers is not clear \citep[e.g.][]{Yungelson1998}. If a remarkable amount of {\it r}-process elements
can be synthesized during AICs \citep{Wheeler1998}, the AIC rate would be constrained to be at least an order of
magnitude lower than $\sim10^4 \rm Gpc^{-3} yr^{-1}$ to be consistent with the observed abundances of neutron-rich
elements in the universe \citep{Fryer1999}.

In any case, by considering of the possible high beaming of FRB radiation (i.e., $f_{\rm b}\ll1$), the inferred extremely high rate of FRBs could be a serious problem for any kinds of CBMs. A possible solution of this problem is that FRBs could actually be produced by the merger products but not by the mergers themselves and, furthermore, the FRBs are all repeated just on a
timescale longer than the period (i.e., several years) of current
observations. If the merger products can produce an FRB averagely on a
timescale of $t_{v}$ during an activity period of $nt_{v}$, then the
rate presented in Equation (\ref{mergerrate}) can be reduced by the
factor of $n$. In this case, a rapidly rotating and highly magnetized NS as a merger product could be most favorable for causing repeatable FRB radiation.
This discussion is applicable for the WD AICs and also for NS-NS mergers. In the latter case, the formation of a massive NS is usually suggested by the afterglow emission of
short GRBs \citep{Dai2006,Fan2006,Rowlinson2013}
and even by the kilonova emission \citep{Yu2017}. In the framework of the merger-produced NS model, the
young NS could power FRBs by its rotational energy as super-giant
radio pulses of pulsars \citep{Connor2016,Cordes2016,Lyutikov2017} or by its magnetic energy as
the giant flares of Galactic magnetars \citep{Popov2010,Kulkarni2014,Katz2016b}. Additionally, a persistent
counterpart associated with the FRBs can be expected to arise from
the interaction of the merger/AIC ejecta with the environmental
materials \citep{Piran2013,Piro2013}. These
characteristics could make regular FRBs similar to the repeated FRB
121102 \citep{Kashiyama2017,Metzger2017,Cao2017b,Dai2017,Michilli2018}, which needs to be
investigated in future.

\acknowledgements
The authors thank Bo Wang and Fa-Yin Wang for valuable discussions. This work is supported by the National Natural Science
Foundation of China (grant nos. 11473008 and 11373006).

\end{document}